%% file: cks.tex
\documentclass[sigconf]{acmart}

\usepackage{balance}
\usepackage{booktabs} \usepackage{enumitem}
\usepackage{booktabs} \usepackage{todonotes}
\usepackage{letltxmacro}
\usepackage{tabularx}
\usepackage{url}
\usepackage{msc}
\usepackage[T1]{fontenc}
\usepackage[utf8]{inputenc}
\PassOptionsToPackage{hyphens}{url}

\LetLtxMacro{\todonote}{\todo}
\renewcommand{\todo}[2][]
{\todonote[inline, caption={#2}, size=\footnotesize, #1]
{\renewcommand{\baselinestretch}{0.5}\selectfont#2\par}}

\usepackage{amssymb}\usepackage{pifont}

\setlist[itemize]{leftmargin=6mm}

\setcopyright{acmlicensed}

\acmConference[SECPID'18]{3rd Workshop on Security, Privacy, and Identity Management in the Cloud}{August 2018}{Hamburg, Germany}
\acmYear{2018}

\ifdefined\abridged
\else

\renewcommand\footnotetextcopyrightpermission[1]{} 
\fi

\begin{document}

\title[Keys in the Clouds]{Keys in the Clouds: Auditable Multi-device \\Access to Cryptographic Credentials}

\author{Arseny Kurnikov}
\affiliation{  \institution{Aalto University, Finland}
}
\email{arseny.kurnikov@aalto.fi}

\author{Andrew Paverd}
\affiliation{  \institution{Aalto University, Finland}
}
\email{andrew.paverd@ieee.org}

\author{Mohammad Mannan}
\affiliation{  \institution{Concordia University, Canada}
}
\email{m.mannan@concordia.ca}

\author{N. Asokan}
\affiliation{  \institution{Aalto University, Finland}
}
\email{asokan@acm.org}

\renewcommand{\shortauthors}{A. Kurnikov et al.}

\begin{abstract}
\input{content/00-abstract}

\end{abstract}

\ccsdesc[500]{Security and privacy~Key management}
\ccsdesc[300]{Security and privacy~Authentication}
\ccsdesc[300]{Security and privacy~Trusted computing}

\keywords{Cloud, Intel SGX, Key management}

\maketitle

\input{content/01-introduction}

\input{content/02-background}

\input{content/03-requirements}

\input{content/04-design}

\input{content/05-implementation}

\input{content/06-evaluation}

\input{content/07-comparison}

\input{content/08-extensions}

\input{content/09-related_work}

\input{content/10-conclusion}

\begin{acks}
This work was supported in part by ICRI-CARS at Aalto University, and the Cloud Security Services (CloSer) project (3881/31/2016) funded by Tekes/Business Finland.
M.~Mannan is supported in part by an NSERC Discovery Grant and a NordSecMob Scholarship.
The authors thank Phil Zimmermann for helpful discussions about this work.
\end{acks}

\balance

{
\raggedright
\bibliographystyle{ACM-Reference-Format}
\bibliography{cks}
}

\end{document}

%% file: content/00-abstract.tex
Personal cryptographic keys are the foundation of many secure services, but storing these keys securely is a challenge, especially if they are used from multiple devices.
Storing keys in a centralized location, like an Internet-accessible server, raises serious security concerns (e.g.\ server compromise).
Hardware-based Trusted Execution Environments (TEEs) are a well-known solution for protecting sensitive data in untrusted environments, and are now becoming available on commodity server platforms.

Although the idea of protecting keys using a server-side TEE is straight-forward, in this paper we validate this approach and show that it enables new desirable functionality.
We describe the design, implementation, and evaluation of a TEE-based Cloud Key Store (CKS), an online service for securely generating, storing, and using personal cryptographic keys.
Using remote attestation, users receive strong assurance about the behaviour of the CKS, and can authenticate themselves using passwords while avoiding typical risks of password-based authentication like password theft or phishing.
In addition, this design allows users to 
i) define policy-based access controls for keys;
ii) delegate keys to other CKS users for a specified time and/or a limited number of uses; and
iii) audit all key usages via a secure audit log. 
We have implemented a proof of concept CKS using Intel SGX and integrated this into GnuPG on Linux and OpenKeychain on Android.
Our CKS implementation performs approximately 6,000 signature operations per second on a single desktop PC.
The latency is in the same order of magnitude as using locally-stored keys, and 20x faster than smart cards.

%% file: content/01-introduction.tex
\section{Introduction}
\label{sec:introduction}

Personal cryptographic keys are the foundation of many secure services, such as signing or decrypting emails, signing code, authenticating to remote servers, or decrypting cloud storage.
However, storing personal cryptographic keys \emph{securely} often proves to be difficult for users, especially when keys must be usable from multiple devices (e.g.\ PCs, smartphones, and tablets).
Although we already have various approaches for protecting personal cryptographic keys, these have various limitations:

\textbf{Password only:} keys stored directly on the user's device may be compromised by malicious software on the device.
Even if the keys are protected by a password, either the password or the decrypted key could be captured by malicious software (see e.g.~\cite{InvisibleThings_keys}).

\textbf{Device key store:} storing keys in e.g.\ the hardware-backed Android or iOS key store protects them from malicious software, but this type of device key store may not be available on all the user's devices.
Furthermore, the keys may be vulnerable while they are being transferred between devices.

\textbf{External peripherals:} smart cards or USB tokens (e.g.\ Trezor~\cite{Trezor} and YubiKey~\cite{Yubikey}) can protect keys from malicious software whilst being usable from multiple devices.
However, they require additional hardware or peripherals (e.g.\ smart card reader), which incur costs and may be incompatible with some devices.

It is also important to consider the \emph{availability} of the keys: keys that only exist on a single device (e.g.\ generated and used exclusively within a hardware-backed device key store or smart card) would become unavailable if the device/smart card is lost or damaged.

Storing personal cryptographic keys in a centralized location (e.g.\ an Internet-connected server) solves many of the above challenges: users can authenticate and use their keys from any Internet-connected device, and centralized servers generally have lower risk of loss or failure than individual users' devices.
However, the principal challenge is how to \emph{protect the keys} against external attackers, other users, and even malicious server administrators.

\textbf{Password-based encryption:} a naive approach is to encrypt keys using the user's password and download the encrypted keys to the user's devices when needed.
However, since passwords are generally weak secrets~\cite{Bonneau2012}, this does not provide sufficient protection against an adversary who can obtain the encrypted keys (e.g.\ via a compromised server) and will eventually be able to guess many users' passwords.
Furthermore, the decrypted keys would still be vulnerable to any malicious software on the user's device.

\textbf{Key splitting:} another approach is to split a key into two (or more) shares, one of which is held by the server and another by the user's device (e.g.~\cite{Cocks1998, MacKenzie2001}).
In order to use the key, both the server and the user's device must cooperate.
Although this approach protects the keys against server compromise, it entails a relatively complicated process in order to use the key from a new device (e.g.\ a share of the key must be securely transferred to the new device).

\textbf{Trusted Execution Environment (TEE):} hardware-enforced TEEs, like Intel SGX~\cite{intel-sgx} and ARM TrustZone~\cite{TrustZone}, can be used to isolate and protect a small amount of trustworthy code and data from all other software on the machine, including the OS and hypervisor.
Furthermore, by using \emph{remote attestation}, a TEE can provide strong assurance to remote parties about precisely what code it is running, and can establish a secure communication channel.
Server-side TEEs are therefore a promising solution for protecting cryptographic keys on a centralized server, and are now becoming available in commodity server hardware~\cite{sgx-servers}.

\ifdefined\abridged
\else
Storing cryptographic keys is a well-known use case for TEEs.
However, in this paper we validate the assertion that server-side TEEs can be used to protect personal cryptographic keys, and we show that such a design can in fact provide several desirable features that could not otherwise be realized.
\fi
We present the design, implementation, and evaluation of a \emph{Cloud Key Store} (CKS) in which keys are generated, stored, and used exclusively within a server-side TEE.
We refer to these as \emph{protected keys}.
Using remote attestation, the user receives strong assurance that she is communicating with a legitimate TEE, and can establish a unilaterally-authenticated secure communication channel directly to the TEE.
The user authenticates herself to the TEE by sending her username and password via this channel (i.e.\ achieving mutual authentication).
Password-based authentication is still the most widespread user authentication method~\cite{Bonneau2012a} and can be used from any of the user's devices.
By rate-limiting the authentication attempts for each username, the TEE prevents password guessing attacks, even from a compromised server.
The user submits requests to the CKS, which performs the requested operations inside the TEE, and returns the results via the same channel.

In addition to storing keys securely and providing access from multiple devices, the CKS also enables the following new features:

\begin{itemize}

\item \textbf{Policy-based access control:} the key owner can restrict key use to a specific time period and/or number of uses.

\item \textbf{Key delegation:} users can delegate access to their keys to other users of the same CKS for either a specific time period and/or number of uses.

\item \textbf{Key usage auditing:} every operation performed using the protected keys can be logged by the CKS, and a user can audit these logs to detect any misuse of the protected keys.

\end{itemize}

\noindent
Our contributions are as follows:

\begin{itemize}

\item We design and fully implement a Cloud Key Store (CKS) that supports password-based user authentication, secure key generation, decryption and signature operations, policy-based access control, key delegation, and key usage auditing (Sections~\ref{sec:design} and \ref{sec:implementation}).
Our implementation uses off-the-shelf hardware, and is available as open source software~\cite{project-page}.

\item We demonstrate the functionality of our CKS by integrating it into two applications: GnuPG on Linux and OpenKeychain on Android (Section~\ref{sec:implementation}), which we also provide as open source software~\cite{project-page}.
The core CKS operations do not require any changes in the applications' UI.
The new CKS features (i.e.\ delegation and auditing) are provided via a web interface, but in future could also be integrated into the applications.

\item We evaluate the security and performance of our CKS implementation, using both GnuPG on Linux and OpenKeychain on Android (Section~\ref{sec:evaluation}).
On a single desktop CPU, our CKS implementation can perform approximately 6,000 signature operations per second. 
The average latency is in the same order of magnitude as using an unprotected key on the local device, and 20x faster than using a smart card. 

\item Finally, we present a comparison of different approaches for managing personal cryptographic keys in terms of their security and functionality (Section~\ref{sec:comparison}).
This shows that a CKS outperforms alternative approaches in multiple aspects, whilst providing additional desirable features.

\end{itemize}

%% file: content/02-background.tex
\section{Background}
\label{sec:background}

\subsection{Intel SGX}
Intel Software Guard Extensions (SGX)~\cite{intel-sgx} is one recent instantiation of a TEE.
SGX is a set of CPU extensions that allow applications to define one or more \emph{enclaves} that are protected from all other software on the platform, including the operating system (OS) and hypervisor.
Data within an enclave can only be accessed from code running within the enclave.
The CPU automatically encrypts enclave memory before it leaves the boundary of the CPU package (e.g.\ is written to DRAM)~\cite{Gueron2016}.
This protects the confidentiality and integrity of the enclave's data, even against an adversary with access to the platform hardware.
An untrusted application can invoke functions within an enclave through pre-defined \texttt{ECALLs}.
Enclaves can call functions from the untrusted application via \texttt{OCALLs}.

When an enclave needs to persist data, it can encrypt the data using a CPU-protected key, called a \textit{sealing key} exclusively available to that specific enclave.
This \emph{sealed data} can be safely stored outside the enclave.
SGX provides \textit{monotonic counters} to prevent roll-back attacks on sealed data.
Finally, SGX provides \emph{remote attestation}, a process through which a remote relying party (the \emph{verifier}) can identify the precise code running inside an enclave (the \emph{prover}), and can establish a secure channel directly to the enclave.
Since well-designed enclaves contain only the minimal amount of code necessary for their intended functionality, this code can be audited or otherwise analysed by the verifier (or a third-party of the verifier's choosing) in order to determine that the code is trustworthy.

\subsection{Pretty Good Privacy (PGP)}

Pretty Good Privacy (PGP)~\cite{Zimmermann1994}, and the associated OpenPGP standard (RFC4880)~\cite{rfc4880}, provide a set of algorithms and procedures to encrypt and/or authenticate data.
OpenPGP is the most widely used standard for encrypting and signing e-mail communications.
\ifdefined\abridged
\else
For example, Alice can use PGP to send encrypted messages to Bob, thus protecting the confidentiality of the messages, and/or can sign her messages to assure Bob of the integrity and authenticity of the messages.
\fi

A major challenge in PGP is to establish a trustworthy mapping between public keys and real-world users.
This is usually achieved using a PKI or a web of trust: a reputation system in which trusted users vouch for the public keys of other users.
Since building up trust through this type of system takes time, revoking a key and re-establishing trust in a new key incurs a relatively high cost.

If an attacker gets access to the user's private key, he can decrypt past e-mails sent to the user, and impersonate the user by generating falsified signatures.
The only way to prevent further damage is to revoke the key.
For this reason, users are recommended to instead use subkeys for day-to-day operations, since individual subkeys can be revoked without affecting the main user identity.

GnuPG~\cite{GPG} is an open-source implementation of the OpenPGP standard.
By default, GnuPG stores encrypted private keys on the user's local device, but it can also use keys stored on a smart card or similar device.

%% file: content/03-requirements.tex
\section{System Model and Requirements}
\label{sec:requirements}

\subsection{System Model}

\begin{figure}[t]
    \begin{center}
    \includegraphics[width=1\columnwidth]{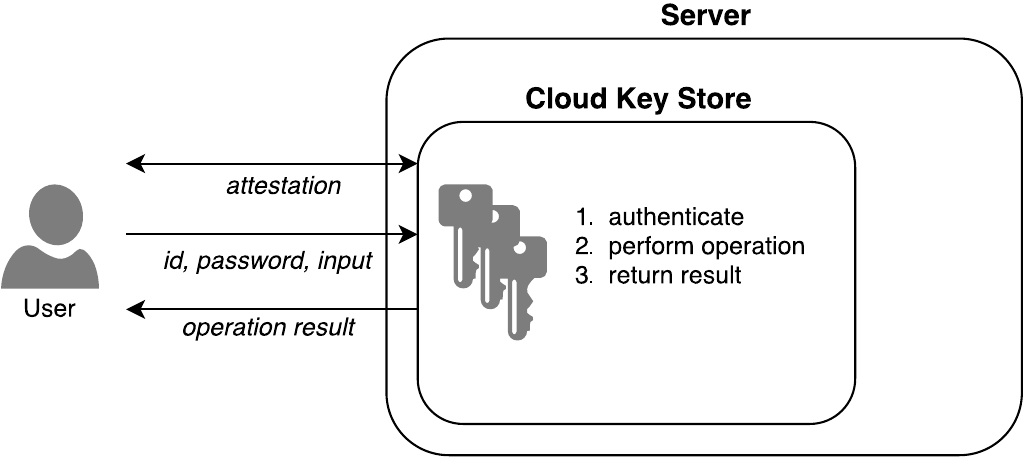}
    \caption{Cloud Key Store architecture.}
    \label{fig:arch}
    \end{center}
\end{figure}

\noindent
Figure~\ref{fig:arch} shows the abstract system model of our CKS running in a Trusted Execution Environment (TEE) on a remote server.
As we discuss in Section~\ref{sec:extensions}, the CKS can also be distributed across multiple physical or virtual servers.

\textbf{Core functionality:} the CKS has the ability to generate, store, and use \emph{protected keys} within the TEE.
The user first runs a \emph{remote attestation} protocol to ensure that the CKS is running in a genuine TEE, and to establish a secure channel to the TEE.
Via the secure channel, the user submits requests to the CKS, containing her username, password, and the inputs for the cryptographic operation (e.g.\ data to be decrypted or signed).
Upon receiving a request, the CKS checks the user's password, performs the requested operation, and returns the result via the secure channel.

\textbf{Additional functionality:} the CKS allows users to delegate usage rights for their keys to other users for a specific period of time and/or number of uses of the key, and to audit all operations that have been performed using their protected keys.

\subsection{Adversary model}
\label{sec:adversary_model}

We assume a strong adversary who has the ability to run arbitrary code on the server with root privileges.
In addition, the adversary has full control of the network, including the ability to monitor, drop, modify, and/or replay any network communication.
We assume the adversary cannot feasibly break correctly implemented cryptographic primitives.
We also assume that the adversary cannot feasibly subvert the security guarantees of hardware-based Trusted Execution Environments (although we discuss side-channel attacks against TEEs in Section~\ref{sec:evaluation}).
In reality, a well-resourced adversary may be able to subvert a hardware-based TEE, but this requires direct physical access and would incur significant cost.
Therefore, our adversary model captures various possible scenarios, including:  

\begin{itemize}
\item An attack and elevation of privileges against the server;
\item A server operator being coerced by law enforcement;
\item A malicious employee of server operator (insider threat).
\end{itemize}

\noindent
In this work, we assume that users do not choose particularly weak passwords (e.g.\ passwords are not in the list of the top 100 most common passwords).
This is a reasonable assumption because i) an average password is estimated to provide approximately 20~bits of security~\cite{Bonneau2012}, and ii) users who are using personal cryptographic keys are likely to be able to choose a password of reasonable strength.

\subsection{Security requirements}
\label{sec:sec-requirements}

The primary security goal of the CKS is to protect keys against exfiltration and/or unauthorized use.
Since key use is authorized by the user's password, the CKS must also ensure the security of this password.
Given the adversary model above, we define the following security requirements to achieve this goal:

\begin{enumerate}[label={R\theenumi},leftmargin=*,labelindent=0mm,labelsep=2mm]

\item \label{R1} \textbf{Authentication:} Protected keys can only be used or delegated by supplying the correct password.

\item \label{R2} \textbf{Offline attacks:} Offline guessing of the password and/or the protected key must be infeasible.

\item \label{R3} \textbf{Online attacks:} Online guessing of the password must be rate limited. Online guessing of protected keys must be infeasible.

\end{enumerate}

\subsection{Performance Goals}
\label{sec:perf_goals}

Ideally, the CKS should not cause a noticeable slow-down from the perspective of the user (e.g.\ compared to using a smart card), and should maximize the throughput per server, thus minimizing the number of servers required.
We therefore define the following two performance goals:

\begin{itemize}

\item \textbf{Latency:} The time required to perform a cryptographic operation using a protected key must be comparable or better than that of using a key stored on a smart card.

\item \textbf{Throughput:} The design should maximize the rate of cryptographic operations performed by the server.
\end{itemize}

\subsection{Deployability Goals}
\label{sec:dep_goals}

The ability to create and restore backups of the CKS is critical for achieving reliable key storage.
This would mitigate the impact of failure and/or theft of the physical server hardware.
The ability to scale horizontally is also necessary to ensure the availability of the CKS, even during periods of peak demand.
It should be possible to dynamically increase/decrease the number of CKS servers on demand.
However, implemented naively, both of these capabilities could undermine security requirement~\ref{R3}, because the adversary could abuse them to increase his rate of password guessing.
We therefore define the following deployability goals:

\begin{itemize}

\item \textbf{Backup and recovery:} The server should support secure backup and recovery procedures to mitigate the risk of hardware failures.

\item \textbf{Scalability:} The server should support secure yet dynamic horizontal scalability (e.g.\ it should be load balancer friendly).

\end{itemize}

\noindent

%% file: content/04-design.tex
\begin{table}[!t]
        \caption{Cloud Key Store operations}
    \label{tab:cks-internal}
    \centering
    \begin{tabularx}{87mm}{|X|l|l|}
            \hline
        \textbf{Name} & \textbf{Input} & \bfseries \textbf{Output} \\
        \hline\hline
        create\_user & uid, new\_pswd, new\_reset\_pswd & -- \\
        pswd\_reset & uid, reset\_pswd, new\_pswd & -- \\
        gen\_key & uid, pswd, key\_attributes & key\_id \\
        import\_key & uid, pswd, key\_data & key\_id \\
        pksign & uid, pswd, key\_id, msg\_hash & signature \\
        pkdecrypt & uid, pswd, key\_id, enc\_msg & decryption \\
        set\_policy & uid, pswd, key\_id, key\_policy & -- \\
        delegate & uid, pswd, key\_id, delegatees, policy & -- \\
        undelegate & uid, pswd, key\_id, delegatees & -- \\
        audit & uid, pswd, key\_id, time\_period & key\_log \\
        delete\_key & uid, pswd, key\_id & -- \\
        \hline
    \end{tabularx}
\end{table}

\section{Design}
\label{sec:design}

\begin{figure}[t]
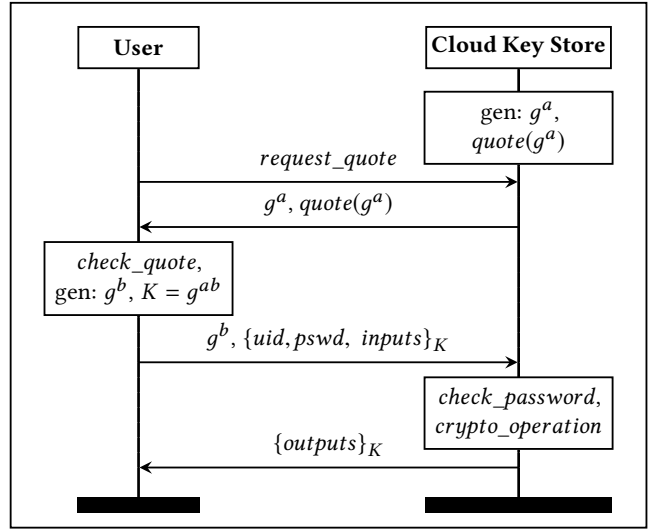


\setlength{\firstlevelheight}{3mm}

\setmsckeyword{}

\begin{msc}{}

\setlength{\instdist}{30mm}
\setlength{\envinstdist}{17mm}
\setlength{\topheaddist}{2mm}
\setlength{\actionwidth}{25mm}
\setlength{\levelheight}{2mm}
\setlength{\bottomfootdist}{2mm}

    \declinst{u}{}{\textbf{User}}
    \declinst{cks}{}{\textbf{Cloud Key Store}}
        \action{gen: $g^a$, $quote(g^a)$}{cks}
    \nextlevel
    \nextlevel
    \nextlevel
    \nextlevel
    \nextlevel
    \nextlevel
    \mess{$request\_quote$}{u}{cks}
    \nextlevel
    \nextlevel
    \nextlevel
    \mess{$g^a$, $quote(g^a)$}{cks}{u}
    \nextlevel
    \action{$check\_quote$,\\ gen: $g^b$, $K=g^{ab}$}{u}
    \nextlevel
    \nextlevel
    \nextlevel
    \nextlevel
    \nextlevel
    \nextlevel
    \nextlevel
    \nextlevel
    \mess{$g^b$, $\left \{uid, pswd,\; inputs\right\}_K$}{u}{cks}
    \nextlevel
    \action{$check\_password$, $crypto\_operation$}{cks}
    \nextlevel
    \nextlevel
    \nextlevel
    \nextlevel
    \nextlevel
    \nextlevel
    \mess{$\left\{outputs\right\}_K$}{cks}{u}
\end{msc}
\caption{Sequence of interactions between the user and the Cloud Key Store.}
\label{fig:msc}
\end{figure}

In this section we present the design of the Cloud Key Store (CKS), focusing on the challenges and our solutions.
The operations that can be performed by the CKS are listed in Table~\ref{tab:cks-internal}, and an overview of the interaction between the user and the CKS is depicted in Figure~\ref{fig:msc}.
In reality, the human user would use some type of client-side software to interact with the CKS (e.g.\ GnuPG on Linux or OpenKeychain on Android).
For clarity of explanation and without loss of generality, we simply use the term \emph{user} to refer to the human user and all client-side software.

\subsection{Establishing a Secure Channel}
\label{sec:design_channel}

To establish trust in the CKS, the user must receive strong assurance that the CKS is running in a genuine TEE, and that only the CKS software is running in the TEE.
This is achieved through remote attestation.
Since we only require unilateral authentication of the TEE towards the user, we use the efficient remote attestation protocol proposed by the SafeKeeper system~\cite{Krawiecka2018}.
This allows the user to verify the TEE, establish a secure channel, and issue a request all in a single message round-trip.
In this section we illustrate the protocol using a discrete logarithm Diffie-Hellman (DH) key agreement protocol, but in practice any equivalent key agreement protocol could be used (e.g.\ Elliptic curve DH). 

Before any users connect, the TEE generates a private key $a$ and a DH public key $g^a$, where $g$ is a generator for a suitable DH group.
All operations are performed in this group (i.e.\ $\mod n$, where $n$ is the group size).
Using the platform functionality, the TEE generates a \emph{quote} containing this public key and full details of the software running in the TEE.
This provides assurance that only the specified software has access to the private key corresponding to $g^a$. 

When a user first connects to the CKS, she requests this quote.
The user verifies the authenticity of the quote, and that the public key supplied by the CKS (or a hash thereof) is included in the quote.
The user then generates her own private key $b$ and DH public key $g^b$, and sends these to the CKS.
At this point the user has established a shared secret key $K=g^{ab}$ with the CKS, and unilaterally authenticated the CKS.
The user encrypts subsequent requests to the enclave with the derived key.

\subsection{Authenticating the User}
\label{sec:design_authenticating}

Once the secure channel has been established, the communication between the user and CKS follows a typical request-response model.
When a request is received, the CKS decrypts the request, looks up the username, and compares the supplied password.
To protect against online password-guessing attacks, the TEE limits the rate of authentication attempts using an exponential back-off algorithm.
When the user authentication fails, the CKS temporarily disables further authentication attempts for that particular user account for a user-defined duration.
If subsequent authentication attempts also fail, the duration between attempts is increased exponentially.
A successful authentication resets the duration to its initial value.
As discussed in Section~\ref{sec:security_evaluation}, the rate-limiting mechanism in the TEE is primarily intended to prevent a compromised server from performing online guessing attacks against the CKS.
The server operator would also implement their own rate-limiting mechanism outside the TEE to prevent external adversaries from causing a denial of service attack by abusing the CKS rate-limiting mechanism.

\subsection{New Users and Password Recovery}

When a user first accesses the system, she uses the \texttt{create\_user} function and supplies a new username, a password, and a reset password.
The username and password must be included in all subsequent requests.
If the user forgets her password, or accidentally reveals it to an adversary, she can send a \texttt{pswd\_reset} request using her reset password.
The limitation of this mechanism is that it still requires the user to store the reset password securely.
However, since this password is not used frequently, it should be feasible for the user to store it securely, e.g.\ written down in a safe.
Other password recovery mechanisms like email may be possible, but would require the user to trust additional services.

\subsection{Policy-based Access Control}

If the authentication succeeds, the CKS performs the requested operation.
The user can generate a new key or import a key using the \texttt{gen\_key} and \texttt{import\_key} functions respectively.
Every key has an associated \emph{key usage policy}, which defines:

\begin{itemize}
\item the username of the key's owner;
\item what types of operations may be performed using the key;
\item the key expiration time;
\item the remaining number of operations.
\end{itemize} 

\noindent
The user can sign or decrypt data using the \texttt{pksign} and \texttt{pkdecrypt} functions respectively.
Every cryptographic operation is checked against the key's usage policy.
If the policy check succeeds, the CKS performs the cryptographic operation and returns the outputs to the user via the secure channel.

The key owner can always change a key's usage policy using the \texttt{set\_policy} function, but this policy is still checked even for the key owner to prevent accidental use of the key (e.g.\ using a signing key for a decryption operation).
The user can delete any keys she owns using the \texttt{delete\_key} function.

\subsection{Key Delegation}

A user can delegate access to protected keys to other users using the \texttt{delegate} function.
The \emph{delegator} specifies the key to be delegated and the usernames of the \emph{delegatees}, as well as any restrictions on the delegation (i.e.\ duration and/or number of uses).
For each delegatee, the CKS first checks that the restrictions match the main key usage policy (i.e.\ a key cannot be delegated beyond its expiry date).
If this check passes, the CKS adds an additional section to the key's usage policy specifying the username of the delegatee, the permitted operations, the duration of the delegation, and the remaining number of operations.
The CKS may add multiple additional sections if the key is delegated to multiple users.

When the delegatee attempts to use this key, the CKS checks the operation against the additional section of the key's usage policy.
The delegatee cannot modify the key's usage policy.
In our current design, the delegatee also cannot delegate these keys further, however, it would be reasonably straightforward to allow this functionality through our flexible key usage policy.

\subsection{Key Usage Auditing}

The CKS maintains a secure log of all operations performed using protected keys.
Before performing an operation, an entry is added to the log specifying the time\footnote{This requires a TEE that provides trusted time capabilities.} of the operation, the key used, the type of operation, the input values, and the output values. 
The user can obtain the log using the  \texttt{audit} command, which takes a key identifier and time period as parameters, and returns all log entries for that key during the specified time period.
If a user notices operations that she did not perform, she can take remedial actions such as resetting the password.
Assuming the user's recovery password has not been compromised, the user does not need to revoke the key -- changing the main password would be sufficient to prevent the adversary from using the key.
If the adversary has delegated the key, this can be reversed using the \texttt{undelegate} function.

%% file: content/05-implementation.tex
\section{Implementation}
\label{sec:implementation}

We implemented the CKS as an Intel SGX enclave running on the server, and we modified two applications to use our CKS: GnuPG on Linux, and OpenKeychain on Android.
In this section we summarize the main technical details of our implementations, which are available as open source software~\cite{project-page}.

\subsection{Enclave}
The enclave's API consists of 3 ECALLs: \texttt{initialize}, \texttt{process}, and \texttt{shutdown}, as explained in the following subsections.
The enclave maintains a \emph{key database} containing the username and password of each user, all cryptographic keys generated or imported by the users, and the associated metadata like the key usage policies.

\subsubsection{Enclave initialization}

The \texttt{initialize} ECALL causes the enclave to generate a fresh DH key pair, pre-generate the information required for remote attestation, unseal the database of keys, and load this into the enclave's memory.
As explained in Section~\ref{sec:design_channel}, our design uses a more efficient remote attestation protocol than the default 4-message protocol provided in the SGX SDK.
To ensure that it has unsealed the latest version of the database, the enclave compares the version number and version nonce in the database against the value of a hardware monotonic counter.
If there is a mismatch, the enclave raises an error and aborts operation.
Before the enclave begins to process requests, it increments the hardware monotonic counter, to indicate that the state of the database may have changed.
This prevents the adversary from crashing the enclave and attempting to restore even the latest version of the database, since the state of the database may have changed.

\subsubsection{User authentication}
The \texttt{process} ECALL is used for all CKS operations.
Compared to creating separate ECALLs for each CKS operation, this design means that the enclave's interface does not need to be changed, even if new CKS functions are added.
This ECALL takes as input an opaque data buffer, the length of the buffer, and the public DH key of the requesting party.

When the enclave receives a request, it first completes the DH key agreement using the public key supplied by the user, and then decrypts the message.
After decryption, the enclave retrieves the username and password from the map of registered users and compares them to the provided values.
As explained in Section~\ref{sec:design_authenticating}, failed authentication attempts cause the enclave to temporarily disable authentication attempts for a user-defined duration, that is exponentially-increasing with subsequent failed attempts.
If the authentication succeeds, the enclave parses the request to obtain the required operation. 

\subsubsection{Signing and decryption}
Our prototype implementation supports both signing and decryption operations using either RSA or Elliptic Curve cryptography.
For RSA, we support the key size of 3072 bits. We support SHA-256 for hashing. For Elliptic Curve cryptography, we use the curve \texttt{secp256r1}. Other algorithms and key sizes can be easily added to the enclave implementation.

\subsubsection{Delegation and trusted time}
In addition to the username and password of the key owner, the delegation function takes as input the username of the delegatee, and the delegation policy that can specify the duration and/or the maximum number of uses allowed.
To enforce the time-limited delegation, our enclave uses the SGX Trusted Time API.
The \texttt{sgx\_get\_trusted\_time} function returns the current time relative to an arbitrary but fixed reference point~\cite{IntelSGXTrustedTime}.
This reference point is guaranteed to remain unchanged as long as the \texttt{time\_source\_nonce} does not change.
For delegation, there is no need to synchronize SGX and wall-clock times because the enclave measures a time \emph{period} starting from the point of processing the delegation request.
The enclave stores the expected \texttt{time\_source\_nonce} value and checks this whenever the time is checked.
If this value changes, the enclave raises an error and refuses to process requests, since an adversary could have modified the platform's hardware clock.

\subsubsection{Auditing and time synchronization}
Before performing any cryptographic operation the enclave creates an entry in the key usage log.
The log is protected in a similar way to the key database.
Creating the log entry before performing the operation ensures that every operation will be logged.
In the worst case, the adversary could interrupt the enclave between creating the log entry and performing the operation, leaving an additional entry in the log.
This can easily be detected by the user.

When a user wishes to audit the key usage, she requests the log entries for a specified time period.
The enclave scans the log and returns the entries with timestamps in the given time period.
Each log entry contains the timestamp and the key identifier of the key that was used.
In this case, it is desirable to synchronize SGX time to wall-clock time because the log entries contain individual \emph{points} in time.
When the CKS is first created (i.e.\ before any users have been added), the server operator is given a once-off opportunity to specify a time offset value for the enclave.
This is intended to synchronize SGX time to a specific wall-clock time reference point (e.g.\ Unix time).
This is purely a convenience feature and does not have any security implications.
If the server operator provides a different offset, users of this CKS will still see consistent time points relative to a different reference point.

\subsubsection{Enclave shutdown/restart}
The \texttt{shutdown} ECALL causes the enclave to persist all important data in preparation for the enclave to be restarted or the platform rebooted.
The enclave stops processing requests and seals the current state of the key database, including a \emph{version number} and \emph{version nonce} obtained from one of the enclave's hardware monotonic counters.
The sealed database must be provided to the enclave when it is restarted.

\subsubsection{Irrecoverable errors}
In the following situations, the enclave cannot recover without potentially compromising the security guarantees of the system:
\begin{itemize}
\item If the system's hardware clock is changed, the enclave cannot determine whether operations satisfy the key usage policy.
\item If an incorrect version of the sealed database is provided during initialization, the enclave cannot safely use this because an unknown number of changes could have been lost.
\end{itemize}

\noindent
In the above cases, the enclave raises an error and refuses to process any requests.
In a real deployment, the CKS service would be distributed across multiple enclaves, such that failure of any single enclave does not affect the overall availability of the system (see Section~\ref{sec:extensions}).
Note that a compromised server operator can always perform a denial of service attack against the system by simply refusing to process network packets or run the enclave.

\subsection{GnuPG integration}

We integrated our CKS into GnuPG as a replacement for the original smart card daemon.
GnuPG has a modular architecture in which the main process provides the user interface but contacts a GnuPG \emph{agent} for the backend operations.
The agent provides a common interface to perform cryptographic operations and key management, regardless of where the keys are stored.
For example, when using a smart card, GnuPG starts the smart card daemon in the background and uses this daemon to perform the cryptographic operations.
We did not extend the existing smart card daemon to communicate with the CKS, but rather we implemented our own CKS daemon, which provides the same interface as the smart card daemon.
This design choice gives finer control of the daemon operation and will allow us to in future integrate the additional features offered by the CKS (e.g.\ key delegation, auditing) into GnuPG.

All messages exchanged between the users and the CKS are encoded using \emph{S-expressions}~\cite{Mccarthy1960}, a flexible and efficient tree-based data structure.
GnuPG uses S-expressions internally when it utilizes its cryptographic library \texttt{libgcrypt}.
Additionally we implemented an HTTP transport to communicate with the CKS.

GnuPG components communicate with one another via a protocol called Assuan~\cite{Assuan2017}.
This is a simple text-based protocol in which each line corresponds to one request or response.
The first word in the line specifies the command and the arguments follow separated by white-space.
For example, when exporting the key to the smart card, the Assuan request line consists of the keyword \texttt{KEYDATA} followed by the encoded key data.
Our new CKS daemon performs the following operations:
 \begin{enumerate}
    \item When started, request the quote from the remote enclave and verify this using the Intel Attestation Service (IAS).
    \item Generate a new DH key pair and calculate the shared key with the enclave.
    \item Start Assuan server and wait for requests from the GnuPG agent.
    \item Dispatch the commands from the agent to the corresponding command handlers.
    \item If the command is security sensitive, the daemon launches a PIN entry program to request the user's password.
    \item Encrypt the buffer containing the operation and its arguments with the shared key established via remote attestation.
    \item Send the encrypted buffer to the enclave for processing.
    \item Decrypt the enclave's response and pass the result back to the agent.
\end{enumerate}

Our GnuPG integration does not require any changes to the GnuPG UI.
When using a CKS protected key, the user's password is requested in the same way as GnuPG would request the user's PIN when using a smart card.
Thus the core CKS operations (i.e.\ key generation, signing, and decryption) can all be carried out from within GnuPG.
For the additional CKS features (i.e.\ delegation and auditing), we provide a simple client-side web app that communicates with the enclave as described above and allows the user to issue delegation commands and inspect the key usage log.

\subsection{OpenKeychain integration}

Following a similar pattern as for GnuPG, we integrated the CKS into the OpenKeychain~\cite{OpenKeychain} application for Android.
Although there is a version of GnuPG for Android~\cite{GPGAndroid}, this is not being actively maintained, and has not been updated in the past three years.
Therefore we decided to target OpenKeychain.
By default, OpenKeychain assumes that keys will be stored locally on the device, but it also supports smart cards.

Similarly to the integration with GnuPG described above, the smart card functionality of OpenKeychain can be changed to communicate with our CKS, so we omit the details. No changes to the app's UI are required to support the core CKS functionality.

%% file: content/06-evaluation.tex
\section{Evaluation}
\label{sec:evaluation}

\subsection{Security}
\label{sec:security_evaluation}

We analyse the security of our solution based on the requirements defined in Section~\ref{sec:sec-requirements}, and we discuss various other attacks and their mitigations.

\textbf{R1 Authentication:}
By the design of our enclave, the protected keys can only be used if a valid password is provided.
Since the user attests the enclave before she generates or imports keys, she can be sure that it will always perform this authentication step.

\textbf{R2 Offline attacks:}
Whenever the key database leaves the boundary of the enclave, it is sealed using the enclave's sealing key.
This prevents offline attacks against the passwords or stored keys.

\textbf{R3 Online attacks:}
A compromised server can attempt to perform an online password guessing attack against the CKS, by pretending to be a legitimate user.
Correctly guessing a user's password would allow the adversary to use that user's protected keys.
The rate-limiting mechanism (Section~\ref{sec:design_authenticating}) makes it infeasible even for a compromised server to mount a successful online guessing attack.

\textbf{Password compromise:}
Even if the adversary learns the user's password, he cannot exfiltrate keys from the CKS.
He can perform operations using the keys until the user detects it via auditing, but when the key owner (i.e.\ the original user) resets her password, using her recovery password, there is no need to revoke her key, since the adversary will no longer be able to access this key using the password he obtained.
If the adversary has delegated the key to other users, the key owner will see this from the audit log, and can revoke these delegations using the \texttt{delete\_delegation} function.

\textbf{Audit log attacks:}
Since the enclave creates a log entry before performing any operation, the only way for an adversary to corrupt the log is to prevent the enclave from performing the operation after the log entry was created.
However, this only allows the adversary to add spurious events to the log.
The user can easily detect these because he would not receive the result of the requested operation.
However it is not possible for the adversary to perform any operation without entries appearing in the log.

\textbf{SGX side-channel attacks:}
SGX is known to be vulnerable to different side-channel attacks, including memory access pattern attacks~\cite{Xu2015,VanBulck2017,Wang2017sgx}, cache attacks~\cite{Brasser2017,Gotzfried2017}, and branch shadowing attacks~\cite{Lee2017}.
Protecting the user identity is not a security requirement because an adversary can infer it by other means (i.e observing the IP addresses).
The protected keys are not vulnerable to side-channel attacks because we perform all cryptographic operations using the recommended Intel libraries for cryptographic operations, which offer state-of-the-art protection against SGX side-channel attacks.

Normally the identity of the user is only known to the enclave since the username is only included in the encrypted buffer.
However, by monitoring the enclave's memory access pattern, an adversary may be able to infer whether two requests originated from the same user.
Nevertheless, in a real-world deployment, the adversary would probably be able to infer this by other means, such as the source IP addresses of the requests.
Therefore, protecting the user identity is not a security requirement.

\textbf{Denial of service:}
The various protection mechanisms included in our enclave could be used by a compromised server to cause a denial of service attack against the CKS.
However, this attack is not in scope because a compromised server could mount this type of attack in various other ways (e.g.\ dropping network packets or simply refusing to run the enclave).

\subsection{Performance}

We evaluate the performance of the CKS in terms of the throughput and memory consumption of the enclave, and the latency introduced from the user's perspective.
The key generation and import operations are not benchmarked, as they are not executed often, so they do not have a significant impact on the user experience.

\textbf{Enclave throughput:}
To measure the overall throughput of the CKS, we benchmarked our implementation on an Intel Core i5-6500 3.20~GHz CPU with 8GB of RAM running Ubuntu 16.04 and the Intel SGX SDK version 2.1.
Over 10 experiments, the average time required to perform 10,000 signature operations using the \texttt{secp256r1} curve was 1,667~ms~($\pm$91~ms).
This gives an effective rate of approximately 6,000 signature operations per second, although it is likely that server-class CPUs would achieve higher throughput.
Throughput for decryption operations is similar. 

\textbf{Enclave memory:}
The memory footprint depends heavily on the key policies, but a typical user record would consist of 600-700 bytes for an RSA key of length 3072 bits.
So an enclave can serve 100,000 users consuming 100~Megabytes of heap memory.

\textbf{Latency:}
To evaluate the impact on the user experience, we measured the latency of a signing operation with and without the CKS.
We ran GnuPG on an Intel Core i5-6500 3.20~GHz CPU with 8~GB of RAM, running Ubuntu 16.04.
We ran OpenKeychain on a Samsung Galaxy~S6 Edge+ with an Exynos 7420 Octa-core CPU (4x2.1~GHz Cortex-A57 and 4x1.5~GHz Cortex-A53) and 4~GB of RAM, running Android 7.0.
The CKS was run on the same machine described above, and all results are the average over 10 experiments.

As the GnuPG baseline, we measured the time required to create an RSA 3072~bit signature using a key on a GnuPG smart card, attached via a USB card reader.
The average latency was 1,282~ms~($\pm$5~ms).
We measured the same signature operation when using CKS from GnuPG.
To remove the variability of network latency, we ran the GnuPG client and the CKS enclave on the same physical machine communicating over localhost.
The average time required to create the signature was 15~ms ($\pm$1~ms).

As the OpenKeychain baseline, we measured the time required to create an RSA 3072~bit signature using an unprotected key in the device.
The average latency was 37~ms ($\pm$4~ms).
To measure the latency of OpenKeychain using CKS whilst removing the network latency, we connected the Android device to the PC running the CKS enclave via a USB connection.
The average time required to create the signature was 24~ms ($\pm$2~ms).

All of the above measurements intentionally exclude the network latency, since this will depend on the type of network to which the user is connected.
To estimate network latency, we performed an HTTP ping of Amazon Web Service (AWS) regions using the CloudPing tool~\cite{CloudPing}.
On both a wired network connection and a 4G mobile network, the latency to the nearest three AWS regions (Ireland, London, and Frankfurt) was in the order of 50ms.
Assuming a network latency of 50ms, the latency of using CKS is in the same order of magnitude as using a local key on the device, and should not be noticeable to the user.

%% file: content/07-comparison.tex
\section{Key storage comparison}
\label{sec:comparison}

To compare different approaches for securely storing personal cryptographic keys, we develop a comparative evaluation framework that considers aspects of security and functionality.
Using this framework we analysed several common approaches, in comparison with our new CKS approach.
The results are summarized in Table~\ref{table:eval-comparison}.
In this table, a filled circle ($\bullet$) indicates that the approach fully defends against the specified attack or fully provides the respective functionality.
An unfilled circle ($\circ$) means that the approach partially supports the feature, whilst a dash ($-$) shows that the approach is vulnerable to the attack, or does not provide that functionality.

\begin{table}[h!]
\caption{Comparison of key storage approaches}
\label{table:eval-comparison}
    \begin{tabularx}{\columnwidth}{|X|c|c|c|c|c|c|}
    \hline
        & \rotatebox{90}{Password encrypted (local)}
        & \rotatebox{90}{Password encrypted (cloud)}
        & \rotatebox{90}{Smart card}
        & \rotatebox{90}{Local hardware storage}
        & \rotatebox{90}{Hardware token}
        & \rotatebox{90}{Cloud Key Store (CKS)} \\ \hline
       \hline
       \multicolumn{7}{|l|}{\textbf{Security}} \\ \hline
        Secure environment      & -         & -         & $\bullet$ & $\bullet$ & $\bullet$ & $\bullet$ \\ \hline
        Resist offline guessing & -         & -         & $\bullet$ & $\bullet$ & $\bullet$ & $\bullet$ \\ \hline
        Resist online guessing  &           & $\circ$   & $\bullet$ & $\bullet$ & $\bullet$ & $\bullet$ \\ \hline
       \hline
       \multicolumn{7}{|l|}{\textbf{Functionality}} \\ \hline
        No additional hardware  & $\bullet$ & $\bullet$ & -         & $\circ$   & -         & $\bullet$ \\ \hline
        Multi-device use        & -         & $\bullet$ & $\circ$   & -         & $\circ$   & $\bullet$ \\ \hline
        Recovery                & -         & $\circ$   & -         & -         & -         & $\bullet$ \\ \hline
        Auditability            & -         & -         & -         & -         & -         & $\bullet$ \\ \hline
        Time-bounded delegation & -         & -         & $\circ$   & -         & $\circ$   & $\bullet$ \\ \hline
        Use-bounded delegation  & -         & -         & -         & -         & -         & $\bullet$ \\ \hline
\end{tabularx}
\end{table}

As shown in Table~\ref{table:eval-comparison}, the security considerations include: whether the key is used within a \emph{secure environment}; and whether the stored keys are vulnerable to \emph{offline} or \emph{online guessing} attacks (e.g.\ of a password).
The functionality considerations include: whether an approach requires additional hardware; whether the key can be used from \emph{multiple devices}; whether the user can \emph{recover} from loss of a device without having to revoke the key; whether the user can \emph{audit} the use of the key; and whether the user can \emph{delegate} access to the key for either a bounded duration or number of uses.

Note that in some cases, certain functionality could be provided by using application-specific mechanisms.
For example, time-bounded delegation could be achieved for PGP signing keys by generating and signing a subkey for a limited duration.
However, we exclude these application-specific mechanisms from the framework because they cannot be used in all cases (e.g.\ the subkey approach cannot be used for decryption keys).

\textbf{Password encrypted (local):}
Keys stored on the user's local device, even if encrypted with a password, are used directly by the applications, which would not typically be considered a \emph{secure environment}.
These keys are vulnerable to offline guessing attacks against the password (online attacks are not applicable).
In terms of functionality, this approach does not require any additional or specialized hardware, however, by default the key cannot be used from multiple devices, cannot be recovered if lost, and cannot be audited or delegated.

\textbf{Password encrypted (cloud):}
The encrypted key is stored in the cloud and downloaded to users' devices when needed. 
Compared to local storage, the security remains mostly unchanged, except that partial protection against online guessing attacks may be provided if the cloud server is trusted.
In terms of functionality, this allows the key to be used from multiple devices, and provides partial recovery capabilities if the user's device is lost but the password is not leaked.
This approach does not provide centralized auditability since the key is downloaded to the user's device.

\textbf{Smart card:}
As expected, a smart card provides significantly better security guarantees: all operations take place in a secure environment, and the smart card is designed to resist offline and online guessing of the password or PIN.
However, in terms of functionality, smart cards themselves are additional hardware and also require a smart card reader.
A single smart card can be used from multiple devices, provided they all have compatible readers.
In general, the device and smart card must be in the same physical location, so protected keys cannot be used concurrently from two different locations.
Smart cards do not provide any recovery mechanism in case of loss or damage.
To some extent, smart cards can enable time-bounded delegation: the key owner can give the smart card to the delegatee, provided she receives it back at the agreed time.

\textbf{Local hardware storage:}
Keys stored in e.g.\ the Android or iOS device key stores enjoy similar protection to that of a smart card.
Although this approach does not require additional hardware, it is not by default possible to use the keys from multiple devices or delegate them to other users.

\textbf{Hardware token:}
Secure hardware tokens like Trezor~\cite{Trezor} and YubiKey~\cite{Yubikey} are very similar to smart cards.
Although they are by definition additional hardware, they usually do not require readers, like with smart cards.

\textbf{Cloud Key Store}
In comparison, the CKS performs all operations within a secure environment, and resists offline and online guessing attacks, as explained in Section~\ref{sec:security_evaluation}.
By design, the CKS does not require additional hardware, and can be accessed from multiple devices concurrently.
As described in Section~\ref{sec:design}, the CKS allows the key owner to audit all uses of the key and enables the user to recover from device loss without having to revoke the key.
Finally, the CKS allows secure revokable delegation of keys in either a time-limited or use-limited fashion.

%% file: content/08-extensions.tex
\section{Extensions}
\label{sec:extensions}

\subsection{Group keys}
The user identities framework can be extended to include groups of users.
Similarly to delegating keys to other users, the group user could set security policies on the group keys.
The functions to add and remove group members can be implemented in the same way as the current delegation functions.
Flexible per-user policies can be combined with setting up per-group key policies.
Then the use of group keys can also be use-limited and time-limited.

\subsection{Multiple passwords}
The password authentication mechanism can be extended to support multiple passwords for each user.
For example, the user can set up different passwords for each of her devices.
If one of the user's devices is lost, only the password for that device needs to be reset.

\subsection{Multi-factor user authentication}
The user authentication mechanism can also be extended to support multi-factor authentication e.g.\ using the Universal 2nd Factor (U2F) standard~\cite{u2f} or a one-time password (OTP) algorithm~\cite{rfc6238,rfc4226}.
This would help to prevent misuse of a protected key if the user's password is compromised. 
However, users who activate multi-factor authentication would need to carry and use an additional authentication token or OTP generator app.

\subsection{Trusted time server}
In addition to the trusted time synchronization mechanism described in Section~\ref{sec:implementation}, it would be possible to have the enclave contact a trusted Network Time Protocol (NTP) server over a secure channel.
This would allow the enclave to synchronize its local SGX time to wall-clock time from the NTP server.
When attesting the enclave, users could verify that this process took place, and if the users also trust the same NTP server, they can be assured the enclave has been synchronized to the same time reference.

\subsection{Backup, recovery, and scalability}
Since availability is a critical concern, it should be possible to create backups of the CKS key database and restore these on another SGX-enabled machine, to mitigate against hardware failure.
Additionally, in a cloud setting, it is common to replicate a service across multiple servers in order to achieve higher throughput and better load balancing.
Krawiecka et al.~\cite{Krawiecka2018} faced a similar challenge in terms of backup and scalability in the SafeKeeper system.
The solution they proposed was to run the same enclave on multiple physical machines and perform mutual attestation between these enclaves.
The primary enclave can then replicate the sensitive data (i.e.\ the CKS key database) across these different machines, without reducing the security guarantees (since the enclaves are identical).
They proposed a simple consensus protocol that the enclaves could use to agree on which replicas are active and which are backups.
A similar type of approach could be used to provide backup, recovery, and scalability in CKS.

\subsection{User-authorized migration}
In a real-world deployment, the server operator may need to upgrade to a newer version of the CKS enclave.
However, in our design, any changes to the enclave's software will result in a different enclave identity (MRENCLAVE value), and thus the new enclave would not be able to unseal the database of passwords and protected keys.
Allowing the server operator to upgrade the enclave's software without authorization from the users could allow a malicious or coerced operator to surreptitiously `upgrade' to an enclave that does not adequately protect users' keys.
To prevent this attack, users must explicitly authorize any enclave software upgrade.
For example, this could be achieved using a new \texttt{upgrade\_cks} command that authenticates the user via her password (as usual) and includes the identity of the new enclave, which the user has verified as being trustworthy.
With this authorization, the old enclave could attest the new enclave, confirm that its identity matches the user's authorization, and then transfer the user's keys.
This is similar to the credential transfer protocol proposed by Kostiainen et al.~\cite{Kostiainen2011}, but is slightly simpler because we assume both enclaves can be online simultaneously.

%% file: content/09-related_work.tex
\section{Related Work}

The idea of using TEEs for protecting secrets in the cloud has gained significant attention.
Major cloud providers often provide key management services for their users~\cite{AzureKMS, GoogleKMS, AmazonKMS}.
These services can be used to generate, store, and use cryptographic keys, without exposing the keys to potentially untrusted software.
Typically, keys managed by such services are stored and used within a dedicated Hardware Security Module (HSM).
For example, the OpenStack Barbican key management service provides a common API for clients to manage secrets, including passwords, encryption keys, and certificates.
Barbican supports various back-ends, including storing the secrets in an encrypted database or an HSM.
To avoid the performance and scalability limitations the HSM back-end, Chakrabarti et al.~\cite{Barbican2017} developed a new back-end based on Intel SGX.
Similarly, the Fortanix Self-Defending Key Management Service (SDKMS) uses SGX to ensure users that the service provider does not have access to their keys~\cite{Fortanix}.
Both of these SGX-based approaches also use remote attestation to provide assurance to remote verifiers.
However, these solutions are mainly focussed on managing secrets used by cloud customers (e.g.\ secrets used by VMs running on the provider's cloud infrastructure).

In contrast, our Cloud Key Store is designed to be used by end-users directly from their devices, and to provide user-centric capabilities like key usage auditing and key delegation between users.
This is enabled in part by our use of password-based authentication of end-users within the TEE.
Balisane et al.~\cite{Balisane2017} have previously discussed the idea of storing \emph{authentication templates} (e.g.\ password databases) inside a TEE.
However, in their proposed architecture, the server operator would still be able to capture a user-submitted password in transit before it is input to the TEE.
To avoid this, we first establish an end-to-end secure channel between the user and the TEE, by performing a key agreement protocol that is unilaterally authenticated via the TEE's remote attestation, as explained in Section~\ref{sec:design_channel}.
This channel protects the password in transit, even against a compromised server operator.

%% file: content/10-conclusion.tex
\section{Conclusion and future work}

We presented a Cloud Key Store (CKS) -- the system that enables users to use their keys from any device, without additional hardware.
Using Intel SGX, our implementation of the CKS protects users' keys against even a compromised server.
To demonstrate feasibility and evaluate performance, we integrated the CKS into GnuPG on Linux and OpenKeychain on Android.
The CKS achieves the same strong security guarantees as smart cards or hardware tokens, whilst providing enhanced functionality like delegation and auditing.
Our performance evaluation shows that the CKS throughput is high and the latency is in the same order of magnitude as local key storage.
As future work we plan to investigate new additional features that could be realized as part of the CKS.
For example, if a key has been generated within the CKS, the user could be given the ability to provably but temporarily \emph{lock} the key, thus preventing even herself from using the key during the specified time period.
This could be useful from a legal perspective since a user can prove that she will not have access to the key until a specified time point in the future (e.g. preventing decryption of data encrypted under that key until a future date).